\documentclass[12pt]{article}
\usepackage{amssymb,amsmath,epsfig}

\begin{document}

\title{\bf Cardy-Verlinde Formula of Non-Commutative Schwarzschild Black Hole}

\author{G.
Abbas \thanks{Email: ghulamabbas@ciitsahiwal.edu.pk}
\\
Department of Mathematics, COMSATS Institute of \\Information
Technology Sahiwal, Pakistan}
\date{}
\maketitle
\begin{abstract}
Few years ago, Setare \cite{1} has investigated the Cardy-Verlinde
formula of non-commutative black hole obtained by non-commutativity
of coordinates. In this paper, we apply the same procedure to a
non-commutative black hole obtained by the co-ordinate coherent
approch. The Cardy-Verlinde formula is entropy formula of conformal
field theory in an arbitrary dimension. It relates the entropy of
conformal filed theory to its total energy and Casimir energy. In
paper, we have calculated the total energy and Casimir energy of
non-commutative Schwarzschild black hole and have shown that entropy
of non-commutative Schwarzschild black hole horizon can be expressed
in terms of Cardy-Verlinde formula.
\end{abstract}

\

\section{Introduction}

Verlinde \cite{2} proved that the entropy of conformal field theory
in arbitrary dimension is related to its total energy and Casimir
energy, this is known as generalized Verlinde formula (commonly
termed as Cardy-Verlinde formula). Recently, it has been
investigated that this formula hold well for
Reissner-Nordstr$\ddot{o}$m de-Sitter black hole (BH) \cite{3} and
charged Kerr BH \cite{4}. Birmingham and Mokhtari proved the
validity of Birmingham and Mokhtari \cite{5} proved the Verlinde
formula for Taub-Bolt-Anti-de-Sitter BH. Setare and Jamil \cite{6}
discussed the Cardy-Verlinde formula for charged BTZ BH. Many
authors \cite{8}-\cite{13} have proved the validity of
Cardy-Verlinde for different BHs. The purpose of this paper is to
investigate the validity of Cardy-Verlinde entropy formula for NC
Schwarzschild BH.

In classical general relativity (GR), the curvature singularity is
such a point where physical description of the gravitational field
is impossible. This problem can be removed in GR by taking into
account the quantum mechanical treatment to the standard formulation
of GR. Motivated by such reasoning, some BH solutions in
non-commutative (NC) field theory have been derived. In these
solutions, curvature singularity at origin is removed by de-Sitter
core which is introduced due to NC nature of spacetime \cite{14}.
Ansoldi et al. \cite {15} formulated the NC charged BHs solutions,
this was extended to rotating non-commutative BHs case  by Modesto
and Nicollini \cite{16}. Mann and Nicolini \cite{17} have discussed
the cosmological production of NC BHs. The first NC version of
wormholes solution was investigated by Nicolini and Spallucci
\cite{18}. Farook et al. \cite{19} have investigated the higher
dimensional wormhole solutions in NC theory of gravity. Motivated by
such NC correction to BHs, Sharif and Abbas \cite{20} studied the
thin shell collapse in NC Reissner-Nordstr$\ddot{o}$m geometry.
Banerjee and Gangopadhyay \cite{21} derived the Komar energy and
Sammar formula for NC Schwarzschild BH.

Motivated by the recent development in NC theory of gravity, we have
proved that the entropy of NC Schwarzschild BH horizon can be
expressed in terms of Cardy-Verlinde formula. For this purpose, we
have used the Setare and Jamil method \cite{6}. The plan of the
paper is as follows: In section \textbf{2}, we briefly discussed the
the thermodynamical relations of NC Schwarzschild BH and
Cardy-Verlinde formula and proved that entropy of non-commutative
Schwarzschild BH horizon in can be expressed in terms of
Cardy-Verlinde formula. Section \textbf{3} is devoted to the
concluded remarks of the work done.

\section{Non-Commutative Schwarzschild Black Hole and Cardy-Verlinde formula}

According to GR, singularity is such a region of spacetime at which
the usual laws of physics break down. This problem can be removed by
applying the formulation of NC field theory to GR. For example, the
NC BHs are one of the outcomes of string theory. These have such
geometric structure in which curvature singularity is recovered by
the minimal length introduced by the NC nature of coordinates.
Further, all types of NC BHs expose the de-Sitter core due to
quantum fluctuations at the center of the manifold.

The NC formulation of GR is one of the long standing problems which
has no solution yet. An extensive literature survey
\cite{22}-\cite{24}, imply that the application of Moyal
$\star$-product among the tetrad fields in the gravitational action
is a mathematically correct approach but not physically. It is due
to the fact that the expansion of $\star$-product in NC parameter is
truncated upto a desirable order which causes to destroy the
non-local nature of NC theory. This results to face the BH geometry
with the same curvature singularities as in GR. Instead of using
$\star$-product, one can formulate NC form of GR using the
coordinate coherent state approach.

In this approach, the density of point like source in NC spacetime
can be governed by a Gaussian distribution by using the relation
\cite{14}
\begin{equation}\label{1}
\rho=\frac{M e^{-\frac{r^2}{4\Theta}}}{(4\pi \Theta)\frac{3}{2}},
\end{equation}
where $M$ is constant gravitational, $\Theta$ is constant having the
dimension of length squared. The line element for NC Schwarzschild
BH is \cite{14}
\begin{equation}\label{2}
{ds}^2=f(r)dt^2-\frac{1}{f(r)}
dr^2-r^2(d\theta^2+\sin^2{\theta}d\phi^2),
\end{equation}
where $f(r)=1-\frac{4M}{{r}
\sqrt{\pi}}\gamma\left(\frac{3}{2};\frac{r^2}{4\Theta} \right)$ and
$\gamma$ is lower incomplete gamma function which is defined by
\begin{equation}\label{3}
\gamma\left(\frac{a}{b};x \right)=\int^x_0 t^{\frac{a}{b}-1}
e^{-t}dt.
\end{equation}
In the commutative limit
$\frac{r}{\sqrt{\Theta}}\longrightarrow\infty$, i.e.,
${\Theta}\rightarrow0$, Eq.(\ref{2}) reduces to conventional
Schwarzschild metric. The event horizons of BH can be found by
setting $f(r_h)=0$, which yields
\begin{equation}\label{4}
r_h=\frac{4M}{{}
\sqrt{\pi}}\gamma\left(\frac{3}{2};\frac{{r_h}^2}{4\Theta} \right).
\end{equation}
We take the large radius regime $(\frac{{r^2}_h} {4\Theta} >> 1)$
where we can expand the incomplete gamma function to solve $r_h$ by
iteration. Keeping the terms upto order $\sqrt{\Theta
}e^{\frac{-M^2}{\Theta}}$, we find
\begin{equation}\label{5}
r_h\simeq2M\left[1-\frac{2M}{\sqrt{\pi
\Theta}}\left(1+\frac{\Theta}{2M^2}\right)e^{-M^2/\Theta}\right]
\end{equation}
Now the Hawking temperature for NC schwarzschild BH upto order
$\sqrt{\Theta }e^{\frac{-M^2}{\Theta}}$ is given by
\begin{equation}\label{6}
T_H=\frac{1}{8\pi M}\left[1-\frac{4M^3}{\Theta \sqrt{\pi
\Theta}}\left(1-\frac{\Theta}{2M^2}-\frac{\Theta ^2}{4M^4} \right)
e^{-M^2/\Theta}\right].
\end{equation}
The entropy of the NC Schwarzschild BH ($S=A/4=\pi r^2_h$) upto
order $\sqrt{\Theta }e^{\frac{-M^2}{\Theta}}$ is given by
\begin{equation}\label{7}
S={4\pi M^2}\left[1-\frac{4M}{\sqrt{\pi
\Theta}}\left(1+\frac{\Theta}{2M^2}\right)\right]e^{-M^2/\Theta}.
\end{equation}
The generalized form of Cardy formula (also known as Cardy-Verlinde
formula) is given by \cite{6}
\begin{equation}\label{8}
S_{CFT}=\frac{2\pi R}{\sqrt{ab}}\sqrt{E_C(2E-E_c)},
\end{equation}
where $a,b>0$, $R$ is radius of $n+1$ dimensional FRW universe,
$E_C$ is the Casimir energy and $E$ is the total energy of
underlying field. The definition of Casimir energy is derived by the
violation of Euler relation as \cite{8}
\begin{equation}\label{9}
E_C=n(E+PV-TS-\Phi Q-\Omega J),
\end{equation}
where the pressure for CFT is $P=\frac{E}{nV}$, $J$, $Q$ are zero
for NC Schwarzschild BH, $V$ is the volume of the system bounded by
the apparent horizon. The total energy may be written as sum of
extensive part $E_E$ and Casimir energy $E_C$ as
\begin{equation}\label{9a}
E=E_E+\frac{1}{2}E_c,
\end{equation}
The Casimir energy $E_C$ as well as purely extensive part of energy
$E_E$ can be expressed in terms of entropy $S$ and radius $R$,
\begin{eqnarray}\label{10}
&&E_E=\frac{a}{4\pi R}S^{1+\frac{1}{n}},\\\label{11}
&&E_C=\frac{b}{2\pi R}S^{1-\frac{1}{n}}.
\end{eqnarray}

After the work of Witten \cite{25} on the AdS$_d$/CFT$_{d-1}$
correspondence, Savonije and Verlinde \cite {26} proposed that
Cardy-Verlinde formula can be derived using the thermodynamical
relations of arbitrary BHs in arbitrary dimensions. In this point of
view, we shall prove the validity of Cardy-Verlinde formula for NC
Schwarzschild BH.

 From Eqs.(\ref{8}) and (\ref{9a}), we get
\begin{equation}\label{9b}
S_{CFT}=\frac{2\pi R}{\sqrt{ab}}\sqrt{2E_EE_c},
\end{equation}
Using Eqs.(\ref{10}) and (\ref{11}) in above equation,
\begin{equation}\label{9c}
S_{CFT}=S.
\end{equation}
The Casimir energy given by Eq.(\ref{9}) for $n=2$ with
Eqs.(\ref{6}) and (\ref{7}) takes the following
\begin{eqnarray}\label{12a}
E_C&=&3E-2TS,\\\label{12}&=&3E-M\left[1-\frac{4M^3}{\Theta \sqrt{\pi
\Theta}}\left(1-\frac{\Theta}{2M^2}-\frac{\Theta ^2}{4M^4} \right)
e^{-M^2/\Theta}\right]\nonumber\\&\times&\left[1-\frac{4M}{\sqrt{\pi
\Theta}}\left(1+\frac{\Theta}{2M^2}\right)e^{-M^2/\Theta}\right].
\end{eqnarray}
Using above equation in Eq.(\ref{9a}), we get the purely extensive
part of total energy as
\begin{eqnarray}\label{13a}
E_E&=&-\frac{E}{2}+TS\\\label{13}&=&-\frac{E}{2}+\frac{M}{2}\left[1-\frac{4M^3}{\Theta
\sqrt{\pi \Theta}}\left(1-\frac{\Theta}{2M^2}-\frac{\Theta ^2}{4M^4}
\right)
e^{-M^2/\Theta}\right]\nonumber\\&\times&\left[1-\frac{4M}{\sqrt{\pi
\Theta}}\left(1+\frac{\Theta}{2M^2}\right)e^{-M^2/\Theta}\right].
\end{eqnarray}
Further

\begin{eqnarray}\label{14a}
2E-E_C&=&-{E}+2TS\\\label{14}&=&-{E}+{M}\left[1-\frac{4M^3}{\Theta
\sqrt{\pi \Theta}}\left(1-\frac{\Theta}{2M^2}-\frac{\Theta ^2}{4M^4}
\right)
e^{-M^2/\Theta}\right]\nonumber\\&\times&\left[1-\frac{4M}{\sqrt{\pi
\Theta}}\left(1+\frac{\Theta}{2M^2}\right)e^{-M^2/\Theta}\right].
\end{eqnarray}

From the comparison of Eqs.(\ref{11}) and (\ref{12}), we get
\begin{eqnarray}\label{15a}
R&=&
\frac{bS^{{1/2}}}{4\pi}\left(\frac{3}{2}{E}-TS\right)^{-1}\\\label{15}&=&\frac{bM}{\sqrt{4\pi}}\left[1-\frac{4M}{\sqrt{\pi
\Theta}}\left(1+\frac{\Theta}{2M^2}\right)e^{-M^2/\Theta}\right]^{\frac{1}{2}}\left(\frac{3}{2}E-\frac{M}{2}\left[1-\frac{4M^3}{\Theta
\sqrt{\pi \Theta}}\right.\right.\nonumber\\
&\times&\left.\left.\left(1-\frac{\Theta}{2M^2}-\frac{\Theta
^2}{4M^4} \right) e^{-M^2/\Theta}\right]\left[1-\frac{4M}{\sqrt{\pi
\Theta}}\left(1+\frac{\Theta}{2M^2}\right)e^{-M^2/\Theta}\right]\right)^{-1}.\nonumber\\
\end{eqnarray}

Also, the comparison of Eqs.(\ref{10}) and (\ref{13}), gives
\begin{eqnarray}\label{16a}
R&=&
\frac{aS^{{3/2}}}{4\pi}\left(-\frac{1}{2}{E}+TS\right)^{-1}\\\label{16}&=&{4\pi
a M^3}\left[1-\frac{4M}{\sqrt{\pi
\Theta}}\left(1+\frac{\Theta}{2M^2}\right)e^{-M^2/\Theta}\right]^{\frac{3}{2}}\left(-\frac{1}{2}E+\frac{M}{2}\left[1-\frac{4M^3}{\Theta
\sqrt{\pi \Theta}}\right.\right.\nonumber\\
&\times&\left.\left.\left(1-\frac{\Theta}{2M^2}-\frac{\Theta
^2}{4M^4} \right) e^{-M^2/\Theta}\right]\left[1-\frac{4M}{\sqrt{\pi
\Theta}}\left(1+\frac{\Theta}{2M^2}\right)e^{-M^2/\Theta}\right]\right)^{-1}.\nonumber\\
\end{eqnarray}
Taking the product of Eqs.(\ref{15a}) and (\ref{16a}), we get
\begin{eqnarray}\label{17}
R&=& \frac{\sqrt{ab} }{4\pi}
\frac{S}{\sqrt{\left(-\frac{1}{2}{E}+TS\right)\left(\frac{3}{2}{E}-TS\right)}}.
\end{eqnarray}
Using Eqs.(\ref{12}), (\ref{13}) and (\ref{17}) in Eq.(\ref{8}), we
get
\begin{eqnarray}\label{18}
S_{CFT}=S.
\end{eqnarray}
This result shows that the entropy of the NC Schwarzschild BH  can
be expressed in terms of Cardy-Verlinde formula. As the BH geometric
and thermodynamic quantities are evaluated by assuming large-radius
approximations. So, the Cardy-Verlinde formula is valid only for
large BHs.

\section{Out Look}

As a prolongation of the research on BH and gravitational collapse
\cite{14}-\cite{33} in this paper, we derive the entropy formula in
conformal field theory of a $4D$ static spherically symmetric NC
Schwarzschild BH. This NC BH solution is obtained by introducing the
NC effect through a coordinate coherent state approach, which is in
fact the substitution of the point distributions by smeared source
throughout a regular region of linear size. We perform the analysis
by obtaining entropy and temperature, which show a deviation from
their usual relations depending on the NC parameter $\Theta$. We
have proved that the entropy of the NC Schwarzschild BH can be
expressed in terms of Cardy-Verlinde formula. For this purpose, we
have used the approximate of values of incomplete gamma functions
upto the term $\sqrt{\Theta }e^{\frac{-M^2}{\Theta}}$. With the same
order of approximation the entropy and temperature of NC BH horizons
has been calculated. The procedure adopted in this paper has been
already used by Stare and Jamil \cite{6,8}. It would be interesting
to generalize this work for charged and charged rotating NC BHs. The
Cardy-Verlinde formula of charged NC BH \cite{34} is in progress.

\end{document}